\def\tr{{\text{tr}}\,}

\def\sgn{{\text{sgn\,}}}
\def\be{\begin{equation}}
\def\ee{\end{equation}}
\def\bea{\begin{eqnarray}}
\def\eea{\end{eqnarray}}
\def\bse{\begin{subequations}}
\def\ese{\end{subequations}}

\documentclass[prb,twocolumn,showpacs,amsmath,amssymb,eqsecnum]{revtex4}
\usepackage{graphicx}
\usepackage{dcolumn}
\usepackage{bm}
\begin{document}
\title{Localization in an external electric field}
\author{O. Bleibaum}
\affiliation{Department of Physics and Materials Science Institute,
University of Oregon, Eugene OR 97403}
\affiliation{Institut f\"ur Theoretische Physik, Otto-von-Guericke
Universit\"at Magdeburg, 39016 Magdeburg, Germany}
\author{D. Belitz}
\affiliation{Department of Physics and Materials Science Institute,
University of Oregon, Eugene OR 97403}
\begin{abstract}
The impact of an electric field on the electron localization problem
is studied within the framework of a field-theoretic formulation.
The investigation shows that the impact of the electric field on the
localization corrections is governed by the interplay between two time
scales, one set by the electric field, and the other by the phase relaxation
rate. At very low temperatures the scaling of the conductivity is governed by
the electric field. In this regime the conductivity depends logarithmically
on the field, and an arbitrarily small electric field delocalizes the electron
states. At higher temperatures the behavior of the conductivity is governed by
the temperature scaling. In this regime the field has no impact on the
observable leading localization corrections.
\end{abstract}
\pacs{73.50.Fq. 73.20.Fz, 72.15.Lh}
\maketitle
\section{Introduction}
The physics of weakly disordered electron systems has been the subject of
considerable theoretical and experimental interest over the past years.
According to semi-classical physics, the electrons in such systems move
ballistically in between occasional scattering events due to impurities or
phonons, which results in diffusive dynamics. The semiclassical picture has
long been known to be incorrect in one-dimension, where arbitrarily weak
disorder leads to all electronic states being localized.\cite{Mott_1990} A
completely new understanding of transport in two-dimensions was obtained in the
late 1970s by the scaling theory of Abrahams et al.\cite{Abrahams_et_al_1979}
These authors showed that quantum interferences lead to the localization of all
electronic states in two-dimensions as well, no matter how weak the disorder.
Accordingly, all two-dimensional systems are, strictly speaking,
insulators.\cite{interaction_footnote}

The same conclusion was reached by a field-theoretical approach that was
pioneered by Wegner.\cite{Wegner_1979} Using the replica trick to deal with the
quenched disorder, he derived a generalized nonlinear $\sigma$-model whose
coupling constant is the electrical conductivity. Wegner's original theory used
bosonic fields, but later applications of his method derived the same model
starting from a fermionic formulation.\cite{Efetov_Larkin_Khmelnitskii_1980} A
supersymmetric formulation, which avoids the replica trick, was also
given.\cite{Efetov_1982_a,Efetov_1982_b}

A third approach to the localization problem were mode-mode coupling theories.
While the original formulation by G{\"o}tze\cite{Goetze_1978} missed the
interference effects that lead to localization in two-dimensions, a
self-consistent diagrammatic approach by Vollhardt and
W{\"o}lfle,\cite{Vollhardt_Woelfle_1980} as well as an improvement on
G{\"o}tze's original method,\cite{Belitz_Gold_Goetze_1981} yielded results in
agreement with both the scaling theory and the field-theoretic approaches. They
also agree well with numerical simulations and with experiments.

All of the above approaches focus on the electronic diffusion coefficient, and
make a connection to the electrical conductivity by means of an Einstein
relation. By contrast, experimentally the conductivity is measured by applying
an electric field. This raises the question how an externally applied electric
field affects the localization phenomena.

In the literature, there is no clear answer to this question. Exact results are
available for one-dimensional systems
only.\cite{Prigodin_1980,Delyon_Simon_Souillard_1984,Prigodin_Altshuler_1989}
These are characterized by a critical electric field $F_{\text{c}}$ such that
for field strengths $F<F_{\text{c}}$ the states remain (power-law) localized,
while for $F>F_{\text{c}}$ the states are extended.

Unfortunately, these exact calculations for one-dimensional systems do not
reveal how the electric field affects the quantum interference effects. The
impact of an electric field on the localization in two-dimensional systems is
therefore much less clear. It has been argued that a weak homogeneous electric
field has no influence at all on the
localization.\cite{Altshuler_Aronov_Khmelnitskii_1981,Lei_Cai_1990,
Hershfield_Ambegaokar_1986} Other authors have concluded that an arbitrary
small electric field already leads to delocalization.\cite{Kirkpatrick_1986,
Sondhi_et_al_1997} In between these two extremes lies a theory that predicts a
strong modification of the weak-localization corrections by an electric
field,\cite{Tsuzuki_1981,Tsuzuki_1982} and theories in which delocalization
only occurs if the electric field exceeds a critical
value.\cite{Lee_Chu_Castano_1983,Bryksin_Schlegel_Kleinert_1994,
Bleibaum_et_al_1995}.

Experimentally, the situation is not clear either. While early
experiments\cite{Dolan_Osheroff_1979,Bishop_Tsui_Dynes_1980} seemed to find
modifications of the weak-localization corrections by a field, no impact of the
electric field on the localization corrections was observed in Ref.\
\onlinecite{Bergmann_1982}. Similarly, Ref.\ \onlinecite{Ovadyahu_2001}
concluded that an electric field has no impact on the localization corrections.
On the other hand, the scaling argument of Ref.\ \onlinecite{Sondhi_et_al_1997},
which implies that an arbitrary small electric field leads to delocalization,
is often used in the interpretation of experiments, as discussed in Ref.\
\onlinecite{Abrahams_Kravchenko_Sarachik_2001}.

In the present paper we revisit this problem by means of field-theoretic
techniques. We conclude that delocalization does indeed occur for arbitrarily
small fields. However, this effect is not observable at realistic temperatures,
which explains some of the apparent contradictions between theory and
experiment. We also find that the leading contribution to the electric field
scaling, which was assumed in Ref.\ \onlinecite{Sondhi_et_al_1997} to be the
dominant effect, has a zero prefactor, and the actual effect of the electric
field is weaker. This is important for attempts to extract the dynamical
critical exponents from experimental
data.\cite{Abrahams_Kravchenko_Sarachik_2001}

This paper is organized as follows. In Sec.\ \ref{sec:II} we introduce our
model and formulate an effective field theory. This theory is investigated in a
Gaussian approximation in Sec. \ref{sec:III}, which yields a generalized
diffusion equation for density fluctuations. In Sec. \ref{sec:IV} we derive a
generalized nonlinear $\sigma$-model, which is investigated in a one-loop
approximation. In Sec. \ref{sec:V} we discuss our results and their connections
to previous investigations. Some technical points are relegated to an appendix.


\section{Model, and field-theoretic formulation}
\label{sec:II}

We consider the Hamilton operator
\be
H = H_x + V({\bm x})\quad.
\label{eq:2.1}
\ee
Here
\be
H_x = -\frac{\Delta}{2m} + {\bm{F\cdot x}}\quad,
\label{eq:2.2}
\ee
is the Hamilton operator for free particles in the presence of an electric
field ${\bm F}$, with $\Delta$ the Laplace operator and $m$ the electron mass,
and $V({\bm x})$ is a random potential. We use units such that $\hbar$ and the
electron charge are equal to unity. $V$ is characterized by a Gaussian
distribution with zero mean and second moment
\be
\langle V({\bm x})V({\bm x'})\rangle_{\text{dis}} = \frac{1}{2\pi\nu\tau}\
    \delta({\bm x}-{\bm x'})\quad.
\label{eq:2.3}
\ee
Here $\langle\ldots\rangle_{\text{dis}}$ denotes the disorder average,
$\nu$ is the density of states per spin at the Fermi level, and $\tau$
is the single-particle scattering time.
Unless otherwise noted, we consider a two-dimensional system, $d=2$.

In our investigation we focus on configuration averages of retarded and
advanced Green functions and their products. These functions are solutions of
the differential equations
\be
(\pm i\omega + E - H)\,G^{R,A}({\bm x}, {\bm x'}\vert E;\omega)=
\delta({\bm x}-{\bm x'})\quad.
\label{eq:2.4}
\ee
In equilibrium, ${\bm F} = 0$, $E = \mu$ is the chemical potential. For ${\bm
F} \neq 0$, the quantity $E - {\bm F}\cdot{\bm x} \equiv \mu_x$ is the chemical
potential in a local equilibrium approximation. As a boundary condition, we
require that the Green functions vanish at
infinity.\cite{Kirkpatrick_1986,Bleibaum_et_al_1995} The solutions of Eq.
(\ref{eq:2.4}) then are symmetric with respect to an interchange of ${\bm x}$
and ${\bm x'}$,
\bse
\label{eqs:2.5}
\be
G^{R,A}({\bm x},{\bm x'}|E;\omega) = G^{R,A}({\bm x'}, {\bm x}|E;\omega)\quad.
\label{eq:2.5a}
\ee
This symmetry is a direct consequence of the time-reversal invariance, which is
not broken by the electric field.\cite{Altshuler_Aronov_Khmelnitskii_1981}

The electric field does, however, break the translational invariance in real
space. In the absence of the random potential, or after disorder averaging, a
translation by a vector $\bm{a}$ leads to an energy change $\bm{F\cdot a}$. The
configuration averages $\bar{G}^{R,A}$ of the Green functions therefore satisfy
the relationship
\be
\bar{G}^{R,A}({\bm {x+a}}, {\bm {x'+a}}|E;\omega)=
\bar{G}^{R,A}({\bm x}, {\bm x'}|E-{\bm {F\cdot a}};\omega)\quad.
\label{eq:2.5b}
\ee
\ese
The symmetry properties expressed by Eqs.\ (\ref{eqs:2.5}) will be important
later.

We now consider a generating functional for our Green functions. Following
Ref.\ \onlinecite{McKane_Stone_1981}, we define
\be
Z^{R,A}[j_{R,A}]=\int D[\Phi]\ e^{\pm iS^{R,A}/2
       + \int d{\bm x}\,j_{R,A}({\bm x})\Phi({\bm x})}\quad,
\label{eq:2.6}
\ee
where
\be
S^{R,A}=\int d{\bm x}\ \Phi({\bm x})\,(\pm i\omega + E - H)\,
      \Phi({\bm x})\quad,
\label{eq:2.7}
\ee
with $\Phi({\bm x})$ a real scalar field. The retarded and advanced Green
functions are obtained according to the rule
\be
G^{R,A}({\bm x}, {\bm x'}|E;\omega) =
    \frac{\mp i\,\delta^2}{\delta j_{R,A}({\bm x})\,
\delta j_{R,A}({\bm x'})}\biggl\vert_{j=0}\hskip -5pt \ln Z^{R,A}[j_{R,A}]\,.
\label{eq:2.8}
\ee

To calculate the configuration average we use the replica
trick. To this end we consider $2n$ copies of the original generating
functional ($n$ for the retarded degrees of freedom, and $n$ for the
advanced ones), and take the limit $n\to 0$ at the end of the calculation.
The replicated generating functional takes the form
\bea\label{eq:2.9}
(Z^R Z^A)^n &=& \int \prod_{\alpha}\,D[\Phi_{\alpha}]\
   e^{\sum_{\alpha}\int d{\bm x}\,j_{\alpha}({\bm x})\Phi_{\alpha}({\bm x})}
\nonumber\\
&&\hskip -50pt\times e^{\left[\frac{i}{2}\sum_{\alpha}
   \int d{\bm x}\,\Phi_{\alpha}({\bm x})(i\omega_{\alpha} + E - H)\,
     \Lambda_{\alpha}\Phi_{\alpha}({\bm x})\right]}\quad.
\eea
Here $\alpha = -(n-1)\,,\ldots\,,n$, $\Lambda_{\alpha}=-1$ for $\alpha\leq 0$
and $\Lambda_{\alpha} = 1$ for $\alpha > 0$,
$\omega_{\alpha} = \omega\Lambda_{\alpha}$, and $j_{\alpha} = j_A$ for
$\alpha\leq 0$ and $j_{\alpha} = j_R$ for $\alpha > 0$.
Now we calculate the configuration average
$\bar{Z} = \langle(Z^RZ^A)^n\rangle_{\text{dis}}$, which generates the
averaged Green
functions $\bar{G}^{R,A}$. If one performs a Hubbard-Stratonovich
transformation to decouple the resulting term quartic in $\Phi$, and
then integrates out $\Phi$, one finds
\begin{equation}\label{eq:2.10}
{\bar Z}[j=0]=\int D[Q]\ e^{A[Q]}\quad,
\end{equation}
where
\begin{equation}\label{eq:2.11}
A[Q] = \frac{\pi\nu}{8\tau}\int d{\bm x}\
       \tr \left(Q({\bm x})\right)^2 - \frac{1}{2}\,\tr\ln G_Q^{-1}\quad,
\end{equation}
with
\bea\label{eq:2.12}
(G_{Q}^{-1})_{\alpha\alpha'}({\bm x},{\bm x'}\vert E;\omega)
   &=& \bigl[(i\omega_{\alpha} + E - H_x)\,\delta_{\alpha\alpha'}
\nonumber\\
&& \hskip -40pt + \frac{i}{2\tau}\ Q_{\alpha\alpha'}({\bm x})\bigr]\,
        \delta({\bm x}-{\bm{x'}})\quad.
\eea
It is easy to show that $G_Q$, averaged with respect to the action $A$,
and taken in the limit $n\rightarrow 0$, is diagonal in the replica indices,
and equal to $\bar{G}^R$ ($\bar{G}^A$) for $\alpha>0$ ($\alpha<0$).
The $Q$-matrix fields in the Eqs.\ (\ref{eq:2.11}) and (\ref{eq:2.12}) are
real-valued matrix fields which satisfy the relationship
\begin{equation}\label{eq:2.13}
\Lambda_{\alpha} Q_{\alpha\alpha'}({\bm x}) = \Lambda_{\alpha'}
                 Q_{\alpha'\alpha}({\bm x})\quad.
\end{equation}
The Eqs.\ (\ref{eq:2.11}) - (\ref{eq:2.13}) define the effective action which
forms the basis for our further investigations.

\section{Transport and heating in Gaussian approximation}
\label{sec:III}

\subsection{Saddle-point Green functions}
\label{subsec:III.A}

We now look for a saddle-point solution $Q^{\text{SP}}$ of the effective action
$A[Q]$, defined by
\be\label{eq:3.1}
\frac{\delta A[Q]}{\delta Q}\biggr\vert_{Q=Q^{\text{SP}}} = 0\quad.
\ee
The saddle-point values of $Q$ and $G_Q$ are diagonal in the replica indices,
$Q^{\text{SP}}_{\alpha\alpha'} = \delta_{\alpha\alpha'}\,Q_{\alpha}$,
$(G_{Q^{\text{SP}}})_{\alpha\alpha'} = \delta_{\alpha\alpha'}\, G_{\alpha}$, as
are the exact expectation values. From Eqs.\ (\ref{eq:2.11}, \ref{eq:2.12}) we
find
\bea\label{eq:3.2}
Q_{\alpha}({\bm x})&=& \frac{i}{\pi \nu}\
                  G_{\alpha}({\bm x},{\bm x}\vert E;\omega)
\nonumber\\
                               &=& \frac{i}{\pi \nu}\
    G_{\alpha}(0,0\vert E - \bm{F\cdot x};\omega)\quad,
\nonumber\\
&\equiv& {\tilde Q}_{\alpha}(\mu_x)\quad.
\eea
In the second equation we have used the generalized translational invariance
property of the Green function, Eq.\ (\ref{eq:2.5b}), and the notation in the
third equation emphasizes that the saddle-point field depends on ${\bm x}$ and
the direction of ${\bm F}$ only via $\mu_x=E-{\bm F}\cdot{\bm x}$.

In order to obtain an equation for the Green function we use the generalized
translation invariance property (\ref{eq:2.5b}) and the time reversal
invariance. Due to these properties the Green function in the presence of the
electric field can be written in the form\cite{Bleibaum_et_al_1995}
\bea
G_{\alpha}({\bm x}, {\bm x'}|E;\omega) &=& \int\frac{d{\bm k}}{(2\pi)^d}\
   e^{-i{\bm k}\cdot({\bm x}-{\bm x'})}
\nonumber\\
&&\hskip -40pt\times g_{\alpha}({\bm k}\vert E-{\bm F}\cdot ({\bm x}+{\bm
                                                             x'})/2;\omega)
        \quad,
\label{eq:3.3}
\eea
where $g_{\alpha}({\bm k}|E;\omega)$ is a function that depends only
quadratically on the field ${\bm F}$. If we work to linear order in the field,
the function $g_{\alpha}({\bm k}|E;\omega)$ thus reduces to the equilibrium
Green function,
\be
g_{\alpha}({\bm k}|E;\omega) \approx \frac{1}{i\omega_{\alpha}+E-\epsilon_{\bm
k}
                 + \frac{i}{2\tau}{\tilde Q}_{\alpha}(E)}\quad,
\label{eq:3.4}
\ee
where $\epsilon_{\bm k} = {\bm k}^2/(2m)$. The only difference to the situation
in the absence of the field is then that the quantity $\mu_x$, which has
replaced the chemical potential, goes to zero at the classical turning point,
where ${\bm F}\cdot{\bm x} = E$. We will nevertheless consider $|\mu_x|$ large
compared to $1/(2\tau)$ in calculating the integral in Eq.\ (\ref{eq:3.3}),
i.e., we consider only the region far from the classical turning point. In this
approximation we obtain
\be
Q_{\alpha}({\bm x}) = \sgn\omega_{\alpha}\quad.
\label{eq:3.5}
\ee
In writing down Eq.\ (\ref{eq:3.5}) we have ignored the imaginary part of
$Q_{\alpha}$, which only leads to a weakly field-dependent renormalization of
the chemical potential. The saddle-point Green function obtained in this way
agrees with that derived in Ref.\ \onlinecite{Bleibaum_et_al_1995} in a
self-consistent Born-approximation.

We note that the simple structure of the above saddle-point solution is a
result of the constant density of states in our model. If the density of states
were energy dependent, the saddle-point field would pick up an energy and field
dependent contribution from the density of states, which would modify Eq.\
(\ref{eq:3.5}).


\subsection{Gaussian fluctuations}
\label{subsec:III.B}

We now consider the Gaussian fluctuations about our saddle-point solution.
Substituting $Q = Q^{\text{SP}} + \delta Q$ into Eq.\ (\ref{eq:2.11}), we find
for the Gaussian part of the action
\bea
A^{(2)}[\delta Q] &=& \frac{\pi\nu}{8\tau}\int d{\bm x}\,d{\bm y}
   \sum_{\alpha\alpha'}\delta Q_{\alpha\alpha'}({\bm x})\,\biggl[
       \delta({\bm x}-{\bm y})
\nonumber\\
&& \hskip -70pt -\frac{1}{2\pi\nu\tau}\
       G_{\alpha'}({\bm y},{\bm x}\vert E;\omega)\,
    G_{\alpha}({\bm x},{\bm y}\vert E;\omega)\biggr]\,
           \delta Q_{\alpha'\alpha}({\bm y})\ .
\label{eq:3.6}
\eea
The properties of the Gaussian propagators depend on whether both Green
functions in Eq.\ (\ref{eq:3.6}) are retarded or advanced, or whether one is
retarded and the other advanced. In the former case, the propagator is massive,
in the latter, soft. Setting $\omega = \Omega/2$, the soft propagator $P$
satisfies the equation
\bse
\label{eqs:3.7}
\be
\int d{\bm y}\ \Gamma({\bm x},{\bm y}|E;\Omega)\ P({\bm y},{\bm {x'}}|E;\Omega)
   = \delta({\bm x}-{\bm x'})\quad,
\label{eq:3.7a}
\ee
where
\bea
\tau\Gamma({\bm x},{\bm y}|E;\Omega) &=& \delta({\bm x}-{\bm y})
   - \frac{1}{2\pi\nu\tau}\ G_{\alpha>0}({\bm x},{\bm y}|E;\Omega/2)
\nonumber\\
&&\times G_{\alpha'<0}({\bm y},{\bm x}|E;\Omega/2)\quad,
\label{3.7b}
\eea
\ese
is the corresponding soft vertex. $P$ describes the relaxation of the particle
number density.\cite{Vollhardt_Woelfle_1980}

From Eqs.\ (\ref{eqs:2.5}) we obtain the following properties of the vertex
function $\Gamma$,
\bse
\label{eqs:3.8}
\be
\Gamma({\bm x},{\bm x'}\vert E;\Omega) = \Gamma({\bm x'},{\bm x}\vert E;\Omega)
            \quad,
\label{eq:3.8a}
\ee
and
\be
\Gamma({\bm x}+{\bm a},{\bm x'}+{\bm a}|E;\Omega) =
           \Gamma({\bm x},{\bm x'}\vert E-{\bm F}\cdot{\bm a};\Omega)\quad.
\label{eq:3.8b}
\ee
\ese
These relations are important for a gradient expansion of the vertex function.
With their help, Eqs.\ (\ref{eqs:3.7}) can be written, to second order in the
gradient operator and in the classically accessible region,
\be
(\Omega - {\bm\nabla}\cdot D(\mu_x){\bm \nabla})\,
         P({\bm x},{\bm x'}|E;\Omega) = \delta({\bm x}-{\bm x'})\quad.
\label{eq:3.9}
\ee
Here
\be\label{eq:3.10}
D(\mu_x) = \mu_x\tau/m
\ee
is the electron diffusion coefficient in a local equilibrium approximation. The
details of the derivation are given in Appendix \ref{app:A}. We see that,
inside the classically accessible region, the soft propagator is governed by a
generalized diffusion equation, as one would expect from the fact that
physically, $P$ describes the relaxation of density fluctuations. Outside the
classically accessible region, $\Gamma({\bm x}, {\bm x'}\vert
E;\Omega)\approx\delta({\bm x}-{\bm x'})$. Accordingly, only the modes inside
the classically accessible region are generalized diffusion modes, again in
agreement with what one would expect. Equation (\ref{eq:3.9}) is very similar
to the equation derived in Ref.\ \onlinecite{Kirkpatrick_1986} by kinetic
theory methods.

For later reference we note that, in a schematic notation, the
Gaussian propagator $P$ has the form
\be
P = 1/(\Omega - D{\bm\nabla}^2 + (D/E){\bm F}\cdot
           {\bm\nabla})\quad.
\label{eq:3.11}
\ee
Accordingly, there is a frequency scale
\be
\Omega^* = D{\bm F}^2/E^2\quad,
\label{eq:3.12}
\ee
or a time scale, $t^* = 1/\Omega^*$, that separates diffusive behavior
at $\Omega \gg \Omega^*$ from field-dominated drift behavior at
$\Omega \ll \Omega^*$.

In deriving the generalized diffusion equation we have taken into account
only terms linear in $F$. An estimate of the terms quadratic in $F$ shows
that they are small compared to the ones kept provided that
$\vert{\bm F}\vert\ell/\mu \ll 1$, with $\ell$ the mean-free path.


\subsection{Density relaxation}
\label{subsec:III.C}

The generalized diffusion equation, Eq.\ (\ref{eq:3.9}), differs from the
ordinary diffusion equation by the real-space dependence of $D$, which produces
a term linear in the gradient. This introduces a new singularity into the
differential equation. As a result, the solution of this equation for long
times differs strongly from the ordinary Gaussian one. Let the field point in
$x$-direction, ${\bm F} = (F,0)$, put ${\bm x} = (x,y)$ and ${\bm x}' =
(x',0)$,\cite{translational_invariance_footnote} and perform a Fourier
transform with respect to the direction transverse to the field,
\be
P(x,k;x'\vert E;\Omega) = \int dy\ e^{iky}\,P({\bm x},{\bm x}'\vert E;\Omega)
                           \quad.
\label{eq:3.13}
\ee
We then find that the solution of Eq.\ (\ref{eq:3.9}) takes the form
\bse
\label{eqs:3.14}
\be
P(x,k;x'|E;\Omega) = {\cal P}(\mu_x,|k|;\mu_{x'}\vert E;\Omega)\quad,
\label{eq:3.14a}
\ee
where
\begin{widetext}
\bea
{\cal P}(\mu,p;\mu'\vert E;\Omega)&=&\frac{1}{D'F}\,e^{-p(\mu + \mu')/F)}\,
   \left[\Theta(\mu-\mu')\,M\left(\frac{1}{2}(1+\Omega/D'Fp),1,
                                                  2p\mu'/F\right)\
U\left(\frac{1}{2}(1+\Omega/D'Fp),1,2p\mu/F\right)\right.
\nonumber\\
&&\hskip -70pt \left. +\Theta(\mu'-\mu)\,
   U\left(\frac{1}{2}(1+\Omega/D'Fp),1,2p\mu'/F\right)\
   M\left(\frac{1}{2}(1+\Omega/D'Fp),1,2p\mu/F\right)\right]\
   \Gamma\left(\frac{1}{2}(1+\Omega/D'Fp)\right)\quad.
\label{eq:3.14b}
\eea
\end{widetext}
\ese
Here $D'=D(E)/E = \tau/m$. The functions $U$ and $M$ in Eq.\ (\ref{eq:3.14b})
are the confluent hypergeometric functions.\cite{Abramowitz_Stegun_1972} In
order to obtain this solution we have required that $P$ vanishes at infinity in
the classically accessible region and that the probability current vanishes at
the turning point, so that the particles can not penetrate into the classically
forbidden region.

To understand the nature of this solution we recall that $P$ describes the
relaxation of the particle number density. Let us consider, at time $t=0$, an
ensemble of particles with energy $E=\mu$, located at $x=0$, and with a
homogeneous density in the direction perpendicular to the electric field.
Accordingly, the initial number density is given by
\be
n_0({\bm x},E)=\frac{2\pi}{L}\,\delta(E-\mu)\,\delta(x)\quad,
\label{eq:3.15}
\ee
where $L$ is the linear dimension of the system in $y$-direction. The evolution
of this initial density is governed by the propagator $P$ for $k=0$. In this
limit ${\cal P}$ takes the form
\begin{widetext}
\be
{\cal P}(\mu,0;\mu'\vert E;\Omega) = \frac{2}{D'F}\ \left[\Theta(\mu-\mu')
   \,K_0\left(\sqrt{\frac{4\Omega\mu}{D'F^2}}\,\right)\,
      I_0\left(\sqrt{\frac{4\Omega\mu'}{D'F^2}}\,\right)
   + \Theta(\mu'-\mu)\,K_0\left(\sqrt{\frac{4\Omega\mu'}{D'F^2}}\,\right)
         \,I_0\left(\sqrt{\frac{4\Omega\mu}{D'F^2}}\,\right)\right]\ .
\label{eq:3.16}
\ee
\end{widetext}
Here $I_0$ and $K_0$ are the modified Bessel functions. The inverse Laplace
transformation yields
\be
P(x,0;x'\vert E; t) = \frac{\exp(-\frac{\mu_x+\mu_{x'}}
                                                      {D'F^2t})}{D'Ft}
\,I_0\left(\frac{2\sqrt{\mu_x\mu_{x'}}}{D'F^2t}\right)\ .
\label{eq:3.17}
\ee
Equation\ (\ref{eq:3.17}) explicitly displays the characteristic time
\be
t^* = E/D'F^2\quad,
\label{eq:3.18}
\ee
which was apparent already in Eq.\ (\ref{eq:3.9}) (see Eqs.\ (\ref{eq:3.11},
\ref{eq:3.12})), and which serves as the boundary between the short-time and
the long-time behavior. In order to determine the structure of the generalized
diffusion propagator for $t\ll t^*$, we use Eq.\ (\ref{eq:3.14a}). For $t\ll
t^*$ the spread of the initial $\delta$-package in $x$-direction is small, and
we can use $Fx/E, Fx'/E \ll 1$, in addition to the asymptotic expansion of
$I_0$ for large arguments. If we expand the exponent with respect to $x-x'$, we
find
\be
P(x,0;x'\vert E;t) = \frac{1}{\sqrt{4\pi D(E)t}}\
                      e^{-(x-x')^2/4D(E)\,t}\quad.
\label{eq:3.19}
\ee
As expected, the relaxation of the initial density perturbation is diffusive in
this time regime. Note that in writing Eq.\ (\ref{eq:3.19}) we have ignored the
first moment of the generalized diffusion propagator, which is nonzero. While
the first moment is crucial for calculating currents, it is irrelevant for our
current discussion.

In the opposite limit, $t\gg t^*$, the behavior is very different. In this case
the width of the particle packet becomes very large and the asymmetry of the
particle packet, which is small initially, is getting considerable. The Bessel
function in Eq.\ (\ref{eq:3.17}) approaches unity for asymptotically long
times, so that
\be
P(x,0;{x'}|E,t) = \frac{1}{D'Ft}\ \exp(-\frac{\mu_x+\mu_{x'}}{D'F^2t})\quad,
\label{eq:3.20}
\ee
for $\sqrt{\mu_x\mu_{x'}}/D'F^2t\ll 1$. For very large $|x|$, which satisfy the
requirement $\sqrt{\mu_x\mu_{x'}}/D'F^2t\gg 1$, we obtain
\be
P(x,0;{x'}|E,t) = \frac{1}{D'Ft}\
              \exp(-\frac{(\sqrt{\mu_x}-\sqrt{\mu_{x'}})^2}{D'F^2t})\quad.
\label{eq:3.21}
\ee
These results show that in this regime the dynamics is no longer diffusive.

If, instead of the distribution function (\ref{eq:3.15}), we consider a delta
pulse in both the longitudinal and transverse directions, we have to
investigate the function $P$ for nonzero values of $k$. We have not been able
to do so exactly, and have resorted to a WKB approximation instead. The results
obtained in this way are in qualitative agreement with the case analyzed above.

\subsection{Transport and heating}
\label{subsec:III.D}

So far we have considered the relaxation of density perturbations. We  now turn
to the questions of transport, and Joule heating. To this end we consider,
instead of Eq.\ (\ref{eq:3.15}), an initial density distribution
\be\label{eq:3.22}
n_0({\bm x},E) = \frac{2\pi N}{V}\,\delta(E - \mu - {\bm F}\cdot {\bm x})\quad.
\ee
Here $N$ is the total particle number, and $V$ is the system volume. Notice
that Eq.\ (\ref{eq:3.22}) describes a uniform number density in the classically
accessible region, since
\be\label{eq:3.23}
n_0({\bm x}) = \int\frac{dE}{2\pi}\ n_0({\bm x},E) = \Theta(\mu_x)\,N/V\quad.
\ee

To see how such a density distribution evolves, we realize that ${\cal P}$,
Eq.\ (\ref{eq:3.17}), gives the probability to find a particle at energy $\mu'$
at time $t$, if it had the energy $\mu$ at time $t=0$. In the limit of long
times, $t\gg t^*$, the mean particle energy therefore increases linearly with
time,
\bea
\mu(t) &=& \frac{1}{F}\int d\mu\ \mu\,{\cal P}(\mu,0;\mu'\vert t)
\nonumber\\
       &=& \mu + D'F^2t\quad.
\label{eq:3.24}
\eea
Notice that $P$ and ${\cal P}$ were normalized with respect to an integration
over $x$. Changing the integration variable to $\mu$ results in the additional
factor $1/F$ in Eq.\ (\ref{eq:3.24}). Equation (\ref{eq:3.24}) shows that the
energy fed into the system increases the kinetic energy of the charge carriers.

According to the generalized diffusion equation, the current density
distribution is given by
\be
{\bm j}({\bm x}, E\vert\Omega) = -D(\mu_x)\nabla n(x,E\vert\Omega)\quad,
\label{eq:3.25}
\ee
where $n({\bm x},E\vert\Omega)$ is to be calculated from $n_0$ by means of the
propagator $P$. The volume averaged current density takes the form
\be\label{eq:3.26}
{\bm j}(\Omega) = \frac{1}{V}\int \frac{dE}{2\pi}\,d{\bm x}\
             {\bm j}({\bm x},E\vert \Omega)\quad.
\ee
For our spatially uniform charge density distribution, $n({\bm x},E\vert\Omega)
= n(\mu_x\vert\Omega)$, and we obtain
\begin{equation}\label{eq:3.27}
{\bm j}(\Omega) = -{\bm F}\int\frac{d\mu}{2\pi}\ \frac{dD(\mu)}{d\mu}\,
             n(\mu\vert\Omega)\quad.
\end{equation}
In our Gaussian approximation, the derivative of the diffusion constant is
independent of $\mu$, and thus can be taken out of the integral. We finally
obtain
\be\label{eq:3.28}
{\bm j}(\Omega) = -\frac{\tau N}{mV}\frac{{\bm F}}{\Omega}\quad.
\ee
The Gaussian theory thus yields an Ohmic current that leads to Joule heating.


\section{Nonlinear $\sigma$-Modell}
\label{sec:IV}

\subsection{Effective action}
\label{subsec:IV.A}

The derivation of a matrix field theory in Sec.\ \ref{sec:III} has proceeded in
analogy to the case without an electric field, and the result was structurally
very similar to the latter. In particular, matrix elements that correspond to
products of retarded and advance degrees of freedom are soft, while those
corresponding to products of two advanced or two retarded degrees of freedom
are massive. The chief difference is that, in the presence of an electric
field, the soft modes in Gaussian approximation are not diffusive, but rather
obey the more complicated differential equation (\ref{eq:3.9}).

It is obvious from these observations that one can derive an effective field
theory for the soft modes by repeating the procedure that leads to a nonlinear
$\sigma$-model in the zero-field
case.\cite{McKane_Stone_1981,Pruisken_Schaefer_1982}. As expected, the only
change is that the Laplace operator in the $\sigma$-model is replaced by the
differential operator from Eq.\ (\ref{eq:3.9}). We thus find
\bea
A_{\text{eff}}[\hat{Q}] &=&- \frac{\pi\nu}{8}\int d{\bm x}\ \tr\Bigl(
   {\hat Q}({\bm x})\left[{\bm\nabla}\cdot D(\mu_x){\bm\nabla}\right]
   {\hat Q}(\bm{x})\Bigr)
\nonumber\\
&& - \frac{\pi\nu}{2}\,\Omega\int d{\bm x}\ \tr\Bigl(\Lambda {\hat Q}({\bm x})
         \Bigr)
\quad.
\label{eq:4.1}
\eea
Here $\Lambda$ is a diagonal matrix whose elements have been given after Eq.\
(\ref{eq:2.9}). The matrix elements of ${\hat Q}$ are elements of the
homogeneous space $O(n,n)/O(n)\times O(n)$, and ${\hat Q}$ is subject to the
constraints
\bse
\label{eqs:4.2}
\bea
\tr{\hat Q}(\bm{x}) &=& 0\quad,
\label{eq:4.2a}\\
{\hat Q}^2({\bm x}) &=& 1\quad.
\label{eq:4.2b}\\
{\hat Q}^{\text T}({\bm x}) &=& \Lambda\,{\hat Q}({\bm x})\,\Lambda\quad.
\label{eq:4.2c}
\eea
\ese
These can be incorporated in a parameterization in terms of $n\times n$
matrices $q$,
\be
{\hat Q} = \left(\begin{array}{cc}
                 \sqrt{1+qq^T}&q\\
                 -q^T&-\sqrt{1+q^Tq}\\
                 \end{array}
           \right)\quad.
\label{eq:4.3}
\ee
Roughly speaking, the field $\hat{Q}$ contains the soft parts of the field $Q$
of the previous section.

An expansion of the action, Eq.\ (\ref{eq:4.1}), to quartic order in $q$ reads
\bse
\label{eqs:4.4}
\be
A_{\text{eff}} = A^{(2)} + A^{(4)}\quad,
\label{eq:4.4a}
\ee
with
\begin{widetext}
\bea
A^{(2)} &=& -\frac{\pi\nu}{4}\int d{\bm x} \sum_{\alpha\alpha'}
   q_{\alpha\alpha'}({\bm x})\,\Bigl(\Omega - {\bm\nabla}\cdot D(\mu_x)
       {\bm\nabla}\Bigr)\, q_{\alpha\alpha'}({\bm x})\quad,
\label{eq:4.4b}\\
A^{(4)} &=& \frac{\pi\nu}{32}\sum_{ \{\alpha_i\} } \int d{\bm x}\ \Bigl[
q_{\alpha_1\alpha_2}({\bm x})\,q_{\alpha_3\alpha_2}({\bm x})\, \Bigl(\Omega -
{\bm\nabla}\cdot D(\mu_x){\bm\nabla}\Bigr)\, q_{\alpha_3\alpha_4}({\bm
x})\,q_{\alpha_1\alpha_4}({\bm x})
\nonumber\\
&&\hskip 75pt +\ q_{\alpha_2\alpha_1}({\bm x})\,q_{\alpha_2\alpha_3}({\bm x})\,
\Bigl(\Omega - {\bm \nabla}\cdot D(\mu_x){\bm\nabla}\Bigr)\,
q_{\alpha_4\alpha_3}({\bm x})\,q_{\alpha_4\alpha_1}({\bm x})\Bigr]\quad.
\label{eq:4.4c}
\eea
\end{widetext}
\ese

\subsection{One-loop theory for the diffusivity}
\label{subsec:IV.B}

The expansion of the effective action in powers of $q$ allows for a systematic
loop expansion. To one-loop order we find the following result for the
diffusivity $D$,
\be
D^{(1)}(\mu_x\vert\Omega) = D(\mu_x)\,\left[1-\frac{1}{\pi\nu}\,
                          P({\bm{x}},{\bm x}\vert E;\Omega)\right]\quad,
\label{eq:4.5}
\ee
which has the same structure as in the absence of the electric field. The only
difference is that $P({\bm x}, {\bm x}\vert E,\Omega)$ is calculated from Eq.\
(\ref{eq:3.14b}), according to the relationship
\be
P({\bm x}, {\bm x}\vert E,\Omega) = \int_0^{\lambda} \frac{dp}{\pi}\
   {\cal P}(\mu_x,p\,;\mu_x\vert E,\Omega)\quad,
\label{eq:4.6}
\ee
where $\lambda$ is an ultraviolet cutoff. In the absence of the electric field
this integral is infrared divergent if $\Omega=0$. In the presence of the
electric field we obtain, for $\Omega = 0$, and in the limit $\lambda\mu_x/F\gg
1$,
\be
D^{(1)}(\mu_x) = D(\mu_x)\,\left[1 - \frac{1}{2\pi^2\nu D(\mu_x)}\,
       \ln(\lambda\mu_x/F)\right]\quad.
\label{eq:4.7}
\ee
The static one-loop diffusivity is thus finite in two-dimensions, indicating
that the electric field destroys the mechanism that gives rise to localization
in $d=2$. This was already obvious from Eq.\ (\ref{eq:3.11}), which is less
infrared divergent than a diffusion propagator.

For $\Omega\gg 1/t^*$, with $t^*$ from Eq.\ (\ref{eq:3.18}), the function $P$
reduces to the conventional diffusion propagator, as discussed in Sec.\
\ref{sec:III}. Consequently, in this regime the corrections to the bare
diffusion coefficient take the same form as in the conventional
weak-localization theory in the absence of the field.

\subsection{Scaling analysis}
\label{subsec:IV.C}

We now perform a scaling analysis of the generalized nonlinear $\sigma$-model.
Our procedure is analogous to the one in Ref.\ \onlinecite{us_fermions_I}. To
this end, we deviate from our restriction to $d=2$ and consider the model in
$d\geq 2$, ignoring the complications that arise from the chemical potential
dependence of the density of states in $d>2$. We first write the effective
action in a schematic form that leaves out everything that is not necessary for
power-counting purposes,
\bea
A_{\text{eff}} &=& \frac{1}{G_2}\int d{\bm x}\ {\bm\nabla}^2\,q^2
   + H\int d{\bm x}\ \Omega\,q^2
\nonumber\\
&& \hskip -30pt + \frac{F}{G_2}\int d{\bm x}\ {\bm x}\,{\bm\nabla}^2\,q^2
   + \frac{1}{G_4}\int d{\bm x}\ {\bm\nabla}^2\,q^4 + \ldots
\label{eq:4.8}
\eea
Here $H\propto\nu$, $1/G_2 \propto \nu\tau$, etc. We now
assign a scale dimension $[L] = -1$ to lengths $L$. For $F=0$, this
action contains two fixed points, namely, a stable Gaussian one that
describes the diffusive phase, and a critical one that describes the
Anderson transition.

\subsubsection{Gaussian fixed point}
\label{subsubsec:IV.C.1}

The Gaussian fixed point that describes diffusion in the absence of an
electric field, $F=0$, is characterized by a scale dimension of the field
$q$ equal to
\bse
\label{eqs:4.9}
\be
[q]_{\text{diff}} = (d-2)/2\quad.
\label{eq:4.9a}
\ee
Frequencies must scale like wavenumbers squared in a diffusive phase,
so we also require
\be
[\Omega]_{\text{diff}} = 2\quad.
\label{eq:4.9b}
\ee
$G_2$ and $H$ are then dimensionless,
\be
[G_2]_{\text{diff}} = [H]_{\text{diff}} = 0 \quad,
\label{eq:4.9c}
\ee
and all higher order terms are
irrelevant, with the least irrelevant couplings having a scale dimension
\be
[u]_{\text{diff}} = -(d-2)\quad,
\label{eq:4.9d}
\ee
\ese%
where $u$ denotes a generic irrelevant operator. $1/G_4$ is an example of such
a least irrelevant operator. Adding the electric field, we see that $F$ is
relevant with respect to the diffusive fixed point,
\be
[F]_{\text{diff}} = 1\quad.
\label{eq:4.10}
\ee
The frequency dependent diffusivity, whose bare value is $D = 1/G_2H$,
therefore obeys a scaling law
\be
D(\Omega,F,u) = D(\Omega\,b^2,F\,b,u\,b^{-(d-2)})\quad.
\label{eq:4.11}
\ee
At $\Omega = 0$, and for small $F$, we conclude that the diffusivity
has the structure
\be
D(F) \propto \text{const.} + F^{d-2}\quad,
\label{eq:4.12}
\ee
and in $d=2$ one expects a logarithmic dependence of $D$ on $F$. This is
in agreement with the explicit perturbative result in the previous
subsection.

For $\Omega\neq 0$, $D$ is a function of $F^2/\Omega$, in agreement with the
explicit Gaussian theory in Sec.\ \ref{sec:III}. There are two scaling regimes.
For small $\Omega$, the scaling of $D$ is governed by the electric field, and
for large $\Omega$, the scaling is governed by the frequency. The crossover
between these two scaling regimes is at a frequency $\Omega = \Omega^*$, Eq.\
(\ref{eq:3.12}). In an experiment, the frequency $\Omega$ is effectively
replaced by $1/\tau_{\phi}$, where $\tau_{\phi}$ is the phase relaxation
time.\cite{Abrahams_Anderson_Ramakrishnan_1980} The weak-localization physics
can therefore only be observed if the temperature is large compared to a
crossover temperature $T^*$, which is given by
\be
1/\tau_{\phi}(T^*) = D{\bm F}^2/E^2\quad.
\label{eq:4.13}
\ee
We will further discuss this condition in Sec.\ \ref{sec:V} below.

\subsubsection{Critical fixed point}
\label{subsubsec:IV.C.2}

We now turn to the critical fixed point that describes an Anderson
transition at $F=0$. Here we choose the field $q$ to be dimensionless,
and the scale dimension of the frequency to be $d$,
\be
[q]_{\text{c}} = 0\quad,\quad[\Omega]_{\text{c}} = d\quad.
\label{eq:4.14}
\ee
The bare scale dimension of $G_2$ is then $[G_2]_{\text{c}} = [G_4]_{\text{c}}
= \ldots \equiv [G]_{\text{c}} = 2-d = -\epsilon$. An explicit
renormalization-group calculation shows that $H$ is not renormalized, while the
renormalized counterpart of $G$, $g$, has a fixed point value $g^* =
O(\epsilon)$. The deviations of $g$ from $g^*$ constitute the relevant operator
at the critical fixed point, whose scale dimension determines the correlation
lenght exponent $\nu$. To one-loop order\cite{Wegner_1979}
\be
\nu = 1/\epsilon + O(1)\quad.
\label{eq:4.15}
\ee
The scale dimension of $F$ is also given by a loop expansion, but the leading
term can again be determined just by power counting,
\be
[F]_{\text{c}} = 1 + O(\epsilon)\quad.
\label{eq:4.16}
\ee
The electric field is thus a relevant operator with respect to the
critical Anderson fixed point. In particular, an arbitrarily weak field
$F$ destroys the usual localization in $d=2$.

The above discussion shows only that the $F=0$ fixed point is unstable, and
does not tell what happens instead. The perturbative result, Eq.\
(\ref{eq:4.7}), suggests that there is a metallic phase in $d=2$ for $F\neq 0$.
We note that the result, Eq.\ (\ref{eq:4.16}), is different from the popular
scaling argument which assumes that $F\xi$, with $\xi$ the correlation length,
represents the critical energy or frequency scale, which yields $[F] =
d+1$.\cite{Sondhi_et_al_1997} We will come back to this discrepancy in Sec.\
\ref{sec:V} below.


\section{Discussion}
\label{sec:V}

In summary, we have used field-theoretic methods to investigate the impact of
an electric field on the localization of noninteracting electrons, mostly in
two-dimensions. We have found that there is a characteristic temperature $T^*$
that separates a regime where the physics is dominated by the field from one
where it is not. For $T<T^{*}$ the physics is dominated by the electric field,
which directly affects the structure of the localization corrections. In this
regime the density relaxation is strongly non-diffusive, and the usual
weak-localization corrections to observables are replaced by logarithmic
dependences on the field. The latter have the same structure as those derived
in Refs.\ \onlinecite{Tsuzuki_1982} and
\onlinecite{Bryksin_Schlegel_Kleinert_1994}. Our treatment shows that the
applicability of these results is restricted to $T<T^*$. For $T>T^*$ the
electric field does not significantly affect the diffusion of a particle
packet. Consequently, the weak-localization corrections to the diffusion
coefficient in this regime are the same as in equilibrium, and independent of
the electric field. In this regime the approaches of Refs.\
\onlinecite{Altshuler_Aronov_Khmelnitskii_1981} and
\onlinecite{Hershfield_Ambegaokar_1986} are well founded.

Let us estimate the value of $T^*$ for parameter values that are representative
of a typical weak-localization experiment. \cite{Kravchenko_Klapwijk_2000} With
$D \approx 14\ {\text{cm}}^2/{\text s}$, $F \approx 1.6\times
10^{-2}\text{eV/cm}$, and $E \approx 0.6\ \text{meV}$, one has
$1/\tau_{\phi}(T^*) \approx 10^4 \text{Hz}$. At low temperatures, $\tau_{\phi}$
is dominated by the electron-electron interaction and inversely proportional to
the temperature,\cite{Altshuler_Aronov_Khmelnitskii_1982} $\tau_{\phi} = c/T$.
For the data of Ref.\ \onlinecite{Kravchenko_Klapwijk_2000} we obtain $c
\approx 10^{11}\mbox{K$^{-1}$s$^{-1}$}$. This yields $T^* \approx 10^{-7}\
\text{K}$. We conclude that the crossover between the field-dominated regime
and the usual weak-localization regime occurs at unobservably low temperatures.
This explains why no field-dominated scaling corrections were observed in the
experiments of Refs.\ \onlinecite{Bergmann_1982} and
\onlinecite{Ovadyahu_2001}, and it shows that these observations are not at
odds with the notion that an arbitrarily weak electric field does indeed
destroy the weak localization.

We now come back to the scale dimension of the field $F$ with respect to the
critical fixed point that describes the metal-insulator transition. As
mentioned in Sec.\ \ref{sec:IV}, a scaling argument given by Sondhi et
al.\cite{Sondhi_et_al_1997} assumes that $F\xi$ scales like the critical energy
scale. For the scale dimension of $F$ this yields $[F]_{\text{c}} = z+1$, with
$z$ the dynamical critical exponent. Schematically, this corresponds to a
critical propagator (in the case of an Anderson transition)
\be
\frac{1}{\Omega + D(\nabla)\nabla^2 + F/\nabla}\quad.
\label{eq:5.1}
\ee
This is not consistent with perturbation theory, however, at least not for the
model studied here. From Eq.\ (\ref{eq:3.11}) we see that the actual critical
propagator has the structure
\be
\frac{1}{\Omega + D(\nabla)\nabla^2 + F D({\nabla})\nabla}\quad.
\label{eq:5.2}
\ee
The difference between these two expressions accounts for the difference
between the present result $[F]_{\text{c}}=1$, and that of Ref.\
\onlinecite{Sondhi_et_al_1997}. The reason for the additional gradient squared
in Eq.\ (\ref{eq:5.2}) compared to Eq.\ (\ref{eq:5.1}) is the Ward
identity\cite{Wegner_1979} that is closely related to particle number
conservation. Equation\ (\ref{eq:5.1}) implies that the electric field breaks
particle number conservation, which it does not. The same point can be made at
the level of a fermionic action,\cite{Efetov_Larkin_Khmelnitskii_1980} which
can be seen as follows. The ${\bm F}\cdot{\bm x}$ term in the Hamiltonian, Eq.\
(\ref{eq:2.2}), corresponds to a term
\be
S_F = \int d{\bm x}\int_0^{1/T}d\tau\ ({\bm F}\cdot{\bm x})\ n({\bm x},\tau)
\label{eq:5.3}
\ee
in the action, with $n$ the electron number density field. The latter
corresponds to $n = \tr Q$ in the $Q$-matrix formulation of the field theory,
and in the nonlinear $\sigma$-model this corresponds to $\tr{\hat Q} = 0$. This
shows that the leading coupling of ${\bm F}$ to the electronic soft modes
vanishes for symmetry reasons. The leading nonvanishing term carries an
additional gradient squared. This is the reason why one cannot obtain the
nonlinear $\sigma$-model for electrons in an electric field by simply replacing
the coupling term in the fermionic action by its $Q$-field counterpart. In
contrast, an external magnetic field {\em does} break a symmetry (viz., spin
rotational invariance) and gives some soft modes (viz., the spin-diffusons) a
mass, and replacing the spin density in the Zeeman term by its corresponding
trace over a $Q$-field gives the correct answer. We conclude that, in general,
one cannot extract values for the dynamical exponent $z$ from experimental data
by assuming $[F] = z+1$, as was done in Ref.\
\onlinecite{Abrahams_Kravchenko_Sarachik_2001}.

At this point we would like to note that our model does not take into
account the electron-electron interaction and the resulting inelastic
collisions. Therefore, our results are only valid for
sample sizes smaller than the energy relaxation length. If inelastic
collisions were taken into account, the electron system should be
described by a distribution function with an effective temperature. If the
energy-transfer rate within the electronic subsystem is larger than
the one  between the electron and the  phonon system, the effective
temperature $T_{\text{eff}}$ would  depend also on the strength of the
applied electric field. Since in this case the temperature $T$ in the
in the phase relaxation time $\tau_{\phi}(T)$ would be replaced
by $T_{\text{eff}}$, the field dependence of the effective
temperature would also be reflected in the logarithmic corrections to the
conductivity, as pointed out in
Ref.\ \onlinecite{Anderson_Abrahams_Ramakrishnan_1979}. This can be
mistaken for
a direct impact of the electric field on the localization corrections
in a regime where there is actually none. However, these effects depend
on the ratio between the energy-transfer rate in the electronic subsystem
and the cooling rate. They are therefore
not universal, but rather dependend on the experimental conditions.

We finally discuss the relation of our theory to some previous work in the
literature. We have already noted that our equation for the crucial Gaussian
propagator, Eq.\ (\ref{eq:3.9}), is very similar to the one obtained by
Kirkpatrick with different methods.\cite{Kirkpatrick_1986} Indeed, our theory
is in many respects a field-theoretic version of his kinetic theory. A
different differential equation for the propagator was used in Refs.\
\onlinecite{Lee_Chu_Castano_1983,Bryksin_Schlegel_Kleinert_1994,
Bleibaum_et_al_1995}, which found that a critical field strength $F_{\text{c}}$
is necessary to delocalize the states at the Fermi energy. Instead of Eq.\
(\ref{eq:3.9}), these authors used
\be\label{eq:5.4}
\left(\Omega - D(E){\bm\nabla}^2 + D'(E)\,{\bm F}\cdot{\bm\nabla}
         \right)\,P({\bm x},{\bm x'}\vert E;\Omega)
= \delta({\bm x}-{\bm x'}),
\ee
with $D'(E) = dD(E)/dE$. The use of this equation was based on the notion that
there is a rapid mechanism for cooling, so that heating processes can be
ignored. The resulting equation violates the properties expressed by Eqs.\
(\ref{eqs:3.8}), which are based on time reversal symmetry and generalized
translational invariance. Indeed, if there is a rapid mechanism for cooling
that validates Eq.\ (\ref{eq:5.4}), then the electrons must obviously have
already experienced inelastic collisions before the time at which Eq.\
(\ref{eq:5.4}) becomes valid. They are thus already in the diffusive regime,
$t>t^*$, where the electric field no longer provides the leading effect. In the
opposite regime, $t<t^*$, Eq.\ (\ref{eq:5.4}) is not valid, and Eq.\
(\ref{eq:3.9}) must be used instead.

\acknowledgments We would like to thank Valerij V. Bryksin and Ted Kirkpatrick
for helpful discussions. This work was supported by the DFG under grant No.
Bl456/3-1 and by the NSF under grant No. DMR-01-32555.

\begin{appendix}

\section{Derivation of the generalized diffusion equation}
\label{app:A}

In order to obtain Eq.\ (\ref{eq:3.9}) we have to expand $\Gamma$ about the
$\delta$-function. To this end we need to calculate the moments of $\Gamma$. We
first consider the zeroth moment, which is given in terms of
\be
M_0 = \int d{\bm y}\ { G}_{\alpha}({\bm x},{\bm y}|E;\omega)\,{
G}_{-\alpha}({\bm y},{\bm x}|E;\omega)\quad,
\label{eq:A.1}
\ee
with $\omega = \Omega/2$ and $\alpha>0$. In terms of the function $g_{\alpha}$,
Eq.\ (\ref{eq:3.4}), we have
\bea
M_0&=&\int d{\bm y}\int_{\bm k}\int_{\bm p}\ g_{\alpha}({\bm k}|E -
\bm{F}\cdot(\bm{x} - {\bm y}/2);\omega)
\nonumber\\
&&\hskip -30pt\times  g_{-\alpha}(\bm{p}\,\vert E - \bf{F}\cdot(\bm{x} -
   \bm{y}/2);\omega)\, e^{i(\bm{k} - \bm{p})\cdot\bm{y}}\quad,
\label{eq:A.2}
\eea
where $\int_{\bm k} = \int d{\bm k}/(2\pi)^2$. The terms $\bm{F}\cdot\bm{y}/2$
in Eq.\ (\ref{eq:A.2}) lead to corrections of $O(F^2)$. If we omit them we
obtain
\bea
M_0 &=& \int_{\bm k}\
   g_{\alpha}(K|\mu_x;\omega)\,g_{-\alpha}(K|\mu_x;\omega)
\nonumber\\
&&\hskip -35pt = \nu\int^{\infty}_{-\mu_x}\frac{d\epsilon}{i\omega-\epsilon
                   +\frac{i}{2\tau}\sgn(\omega)}\,
\frac{1}{i\omega-\epsilon_k+\frac{i}{2\tau} \sgn (-\omega)}\quad.
\nonumber\\
\label{eq:A.3}
\eea
In order to evaluate this integral we use the approximation already discussed
in Sec.\ \ref{sec:III}: If $\mu_x>0$ we replace $\mu_x$ by $\infty$, and if
$\mu_x<0$ we replace $\mu_x$ by $-\infty$. In the hydrodynamic limit we then
obtain
\bea
M_0 = 2\pi\nu\tau\,\Theta(\mu_x)\left(1 - \Omega\tau\right)\quad.
\label{eq:A.4}\\
\nonumber
\eea

The first moment ${\bm M}_1$ is defined as
\be
{\bf M}_1 = \int d{\bf y}\ (\bm{x-y})\,{G}_{\alpha}({\bm x},{\bm
y}|E;\omega)\,{G}_{-\alpha}({\bm y},{\bm x}|E;\omega).
\label{eq:A.5}
\ee
In the same approximation as above we find
\bea
{\bm M}_1 &=& \frac{{\bm F}}{4}\frac{d}{dE} \int_{\bm
k}\frac{\partial^2}{\partial{\bm
\kappa}^2}\biggl\vert_{{\bm\kappa}=0}g_{\alpha}({\bm
k}+{\bm\kappa}|\mu_x;\omega\,)
g_{-\alpha}({\bm k}|\mu_x;\omega)\nonumber\\
&=&{\bm{F}}\frac{d}{dE}M_2\quad.
\label{eq:A.6}
\eea
Here $M_2$ is the second moment. It is defined as
\be
M_2 = \frac{1}{4}\int d{\bm y}\ ({\bm y}-{\bm x})^2\,{G}_{\alpha}({\bm
x},{\bm y}|E;\omega)\,{G}_{-\alpha}({\bm y},{\bm x}|E;\omega).
\label{eq:A.7}
\ee
Here we ignore the fact that anisotropic terms can arise due to the electric
field, since any such terms are of $O(F^2)$. Again using the same
approximations as above, we obtain in the hydrodynamic limit
\be
M_2 = \frac{\nu\mu_x}{2m} \int^{\infty}_{-\mu_x}d\epsilon \frac{1}{(1/4\tau^2 +
\epsilon^2)^2} = 2\pi\nu\tau D(\mu_x)\tau\theta(\mu_x).
\label{e:A.8}
\ee
Collecting our results, we now have the following expression for $\Gamma$ in
the hydrodynamic regime, and for ${\bm x}$ in the classically accessible
region,
\be
\Gamma({\bm x},{\bm y}) = \delta({\bm x} - {\bm y})\,\left[\Omega -
   {\bm\nabla}\cdot D(\mu_x){\bm\nabla}\right]\quad.
\label{eq:A.9}
\ee
From this, Eq.\ (\ref{eq:3.9}) follows immediately.
\end{appendix}


\begin{thebibliography}{37}
\bibitem[{tra()}]{translational_invariance_footnote}
\bibinfo{note}{With the field in $x$-direction, one has translational
  invariance in $y$-direction, so this choice can be made without loss of
  generality}.
\end{thebibliography}
\end{document}